# ImmunoAI: Accelerated Antibody Discovery Using Gradient-Boosted Machine Learning with Thermodynamic-Hydrodynamic Descriptors and 3D Geometric Interface Topology

Shawnak Shivakumar
*Department of Computer Science*
Menlo-Atherton High School
Atherton, CA, USA
shawnak.shivakumar@gmail.com

Matthew Sandora
*Department of Chemistry*
Menlo-Atherton High School
Atherton, CA, USA
msandora@seq.org

*Abstract*—Human metapneumovirus (hMPV) poses serious risks to pediatric, elderly, and immunocompromised populations. Traditional antibody discovery pipelines require 10–12 months, limiting their applicability for rapid outbreak response. This project introduces ImmunoAI, a machine learning framework that accelerates antibody discovery by predicting high-affinity candidates using gradient-boosted models trained on thermodynamic, hydrodynamic, and 3D topological interface descriptors. A dataset of 213 antibody–antigen complexes was curated to extract geometric and physicochemical features, and a LightGBM regressor was trained to predict binding affinity with high precision. The model reduced the antibody candidate search space by 89%, and fine-tuning on 117 SARS-CoV-2 binding pairs further reduced Root Mean Square Error (RMSE) from 1.70 to 0.92. In the absence of an experimental structure for the hMPV A2.2 variant, AlphaFold2 was used to predict its 3D structure. The fine-tuned model identified two optimal antibodies with predicted picomolar affinities targeting key mutation sites (G42V and E96K), making them excellent candidates for experimental testing. In summary, ImmunoAI shortens design cycles and enables faster, structure-informed responses to viral outbreaks.

## I. Introduction

Emerging viral threats such as human metapneumovirus (hMPV) pose significant risks to pediatric, elderly, and immunocompromised populations. hMPV has recently caused hundreds of thousands of hospitalizations worldwide, yet no specific antivirals or vaccines are currently available [1]. Traditional monoclonal antibody discovery relies on iterative experimental screens that often take over 10-12 months. Such timelines are ineffective for rapid outbreak response [2]. As an illustrative case, Fig. 1 highlights the sharp rise in hMPV-positive tests associated with the emergent A2.2 lineage during the 2025 respiratory season, underscoring the need for faster therapeutic triage. Computational antibody discovery methods have thus been explored to accelerate this process.

Existing computational approaches fall into two categories. Sequence-based models (e.g., embedding or language models of antibody/antigen sequences) are fast and scalable, but they ignore 3D topological features and often lose binding precision [3]. Physics-based simulations (e.g., molecular dynamics simulations (MD) and free energy perturbation methods (FEP)) can capture detailed interactions, but are time and resource-intensive [4]. These two modeling strategies are illustrated in Fig. 2, which contrasts the sequence-based approach's reliance on learned embeddings with the atomistic resolution offered by physics-based simulations. Additionally, while methods like AlphaFold2 enable rapid structure prediction of antigens, they do not directly predict antibody binding [5].

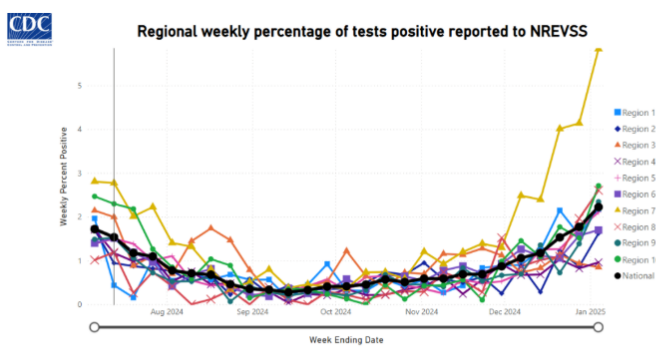

Fig. 1. Surge in hMPV cases reported in January 2025, highlighting the lack of rapid therapeutic discovery tools during viral outbreaks [6].

In this work, we investigate whether structural insights can be integrated into a rapid machine learning framework to triage and predict high-affinity antibody candidates computationally. We hypothesize that a gradient-boosted model trained on detailed 3D interfacial descriptors can efficiently prioritize strong binders.

To this end, we developed ImmunoAI, which, as detailed in Fig. 3: (1) curates a dataset of antibody–antigen complex structures with known affinities, (2) extracts interfacial features (thermodynamic, hydrodynamic, topological), (3) trains a LightGBM model to predict binding affinity, and (4) applies transfer learning on new data (e.g. COVID-19 antibodies, hMPV variants).

Previous work has applied machine learning to antibody affinity prediction, but often relies solely on sequence or ignores spatial detail [7]. In contrast, ImmunoAI leverages structural data and an interpretable ensemble method (LightGBM) to combine speed and accuracy [8]. The key contributions of this study are: a demonstration that 3D-based ML can reduce antibody search space by 89%, transfer learning with SARS-CoV-2 data improves accuracy, and the successful application of the pipeline to predict hMPV-binding antibodies with picomolar affinity, demonstrating the model's potential for guiding real-time therapeutic triage during pandemics.

## II. RELATED WORK

### A. Computational Models

**Sequence-based models:** Recent language models (e.g., ESM-1b, AntiBERTa) embed antibody or antigen sequences and can predict contacts or affinities. These models are fast and can leverage extensive unlabeled data, but they inherently neglect the 3D complementarity of paratopes and epitopes [9].

**Structure generation and prediction**: Tools like RosettaDock and HADDOCK generate structural models of antibody–antigen complexes through docking simulations and rank poses using empirical scoring functions. Separately, AlphaFold2 and AlphaFold-Multimer predict accurate 3D structures from sequence, including multimeric protein complexes like antibody–antigen pairs [5]. However, none of these methods directly compute binding affinities.

**Molecular dynamics simulations**: MD simulates the time-evolution of protein–protein interactions at atomic resolution, enabling extraction of thermodynamic properties such as binding free energy. While accurate, these simulations are computationally intensive. The cost arises from computing pairwise atomic forces and the need to sample long trajectories to achieve convergence. As a result, MD is rarely used for screening large antibody libraries and is more suited for analyzing a few promising binders in detail [4].

### B. Databases and Features

**Databases and prior ML work:** Public databases like Structural Antibody Database (SAbDab) [12] and Antibody Coronavirus (Ab-CoV) [13] provide a curated set of antibody–antigen structures with annotated binding affinities. These have enabled machine learning studies of antibody binding.

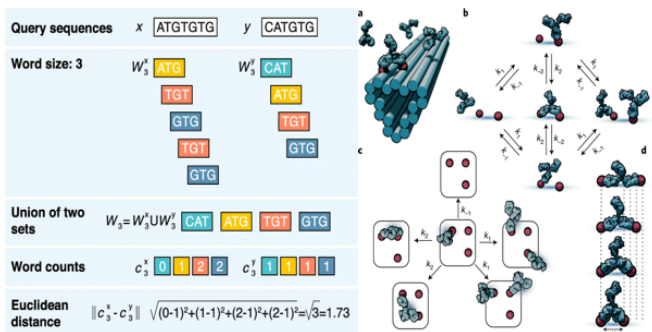

Fig. 2. Conceptual comparison of sequence-based (left) [10] and physics-based (right) [11] antibody triaging approaches.

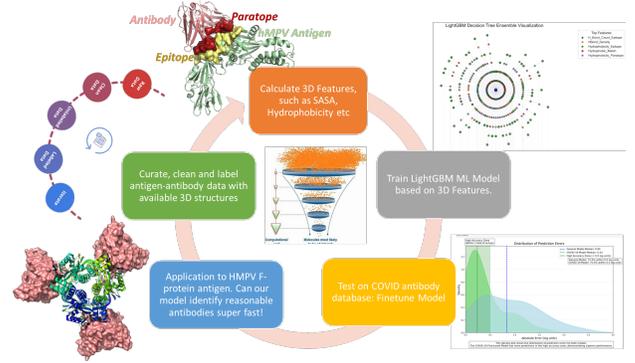

Fig. 3. Methodology pipeline for ImmunoAI: structural feature extraction (SASA, hydrogen bonds, hydrophobicity, etc.) from antibody–antigen complexes, followed by LightGBM affinity prediction and transfer learning.

McElfresh et al. (2023) [14] showed that gradient-boosted tree models often outperform deep nets when data is limited, highlighting the utility of LightGBM in immunoinformatics.

Additionally, decision-tree ensembles like LightGBM offer feature importance insights. In antibody interfaces, features like buried nonpolar surface area and hydrogen bonding networks are known to dominate binding energetics [15]. These insights motivate our focus on physically meaningful 3D descriptors in the ImmunoAI model.

## III. METHODS

Fig. 3 illustrates the full data curation and modeling pipeline: complexes → feature extraction → LightGBM training → fine-tuning on new data → inference on viral mutations.

### A. Data Curation

We curated a training dataset of antibody-antigen complexes with experimentally measured binding affinities. Several databases (AgAbDb, AB-Bind, AACDB, AbDb, etc.) were screened [13]. Ultimately, 213 complexes were selected from the Structural Antibody Database (SAbDab) [12]. Selection criteria included: (i) available Protein Data Bank (PDB) 3D structure, (ii) experimentally measured affinity (in $K_d$), and (iii) human antibodies only (Homo sapiens). Binding affinities in various units (IC50) were converted to dissociation constants (in $K_d$) using the Cheng-Prusoff transformation [16]. The tabular data were split 90% (train) - 10% (test).

### B. Feature Extraction

For each complex, 3D structural features were computed from the interface between antibody and antigen (paratope-epitope) in Table 1. Fig. 6 highlights surface hydrophobicity for an example protein PDB: 6LNU [19].

### C. LightGBM Model Training

The LightGBM framework (an efficient implementation of gradient-boosted decision trees) was chosen for its speed and ability to handle nonlinear feature interactions. The model was trained as a regressor to predict binding affinity (negative log $K_d$) from the structural features.

$$\mathcal{L}^{(m)} = \sum_{i=1}^{n} \left[ g_i h_m(x_i) + \frac{1}{2} h_i h_m(x_i)^2 \right] + \Omega(h_m)$$

Fig. 4. Second-order Taylor expansion of the loss function - incorporates both the gradient *g(i)* and curvature (Hessian) *h(i)* to optimize tree *h(m)* more precisely. Regularization term Ω, *h(m)*, controls tree complexity, which improves generalization in antibody–antigen affinity prediction.

TABLE I. FEATURE SUMMARY (ALL CALCULATED AT INTERSECTION)

| Feature | Description | | |
|---|---|---|---|
| | Type of Descriptor | Calculation | Biophysical Role |
| Solvent-Accessible Surface Area (SASA) | Geometric | Rolling 1.4Å sphere via PyMOL/MSMS | Quantifies buried surface upon binding |
| Hydrophobicity Index | Hydrodynamic | Summed Kyte-Doolittle scores over interface residues | Captures hydrophobic stabilization energy |
| Hydrogen Bonds | Thermodynamic | Counted donor-acceptor pairs ⩽3.5Å | Measures polar interaction strength |
| Solvent-Accessible Energy (SAE) | Thermodynamic | SASA × Hydrophobicity Index | Approximates desolvation and interface energy |
| B-Factor | Topological | Extracted directly from PDB | Indicates atomic mobility and flexibility |
| Atomic Packing Density (APD) | Geometric | Voronoi tessellation with centroid density | Reflects compactness of structural packing |

Table 1. Summary of key structural features used for model input, including descriptor type, calculation method, and biophysical role in antibody–antigen interactions. All features were calculated at the paratope and epitope.

Additionally, the gradient-based residual was utilized to provide direction for model update. This formulation allows the model to dynamically learn from complex biophysical interactions between paratope and epitope features. The second-order approximation improves stability and reduces overfitting compared to simpler boosting strategies.

$$r_i^{(m)} = - \left[ \frac{\partial \mathcal{L}(y_i, \hat{y}_i^{(m-1)})}{\partial \hat{y}_i^{(m-1)}} \right]$$

Fig. 5. Gradient-Based Residual - Negative first-order gradient of loss function; uses partial derivatives to compute pseudo-residuals for boosting.

The LightGBM hyperparameters optimized included:

- Learning Rate: controls the step in gradient descent.
- Number of Leaves: determines the maximum tree complexity (leaf-wise growth).
- Minimum Data in Leaf: samples per leaf to prevent overfitting.

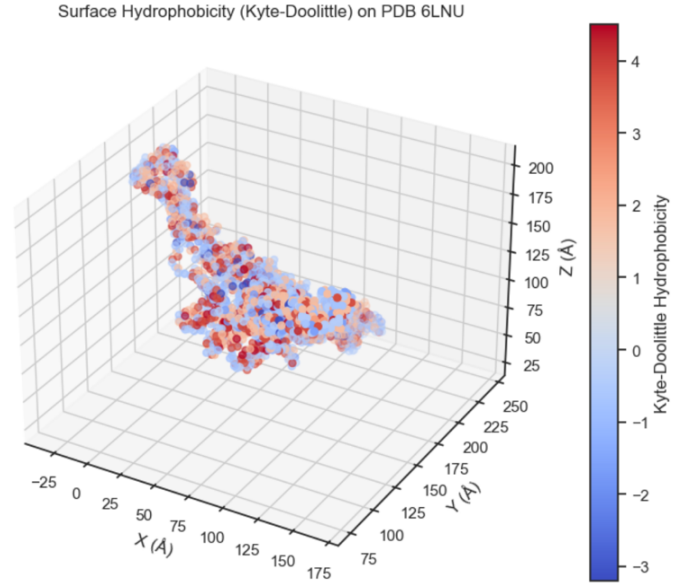

Fig. 6. Surface hydrophobicity map of a protein using the Kyte-Doolittle scale. Red regions indicate hydrophobic surface patches; blue regions indicate hydrophilic areas.

We performed cross-validation on the training set to tune these parameters. The initial baseline model (prior to fine-tuning) was trained on the 213-complex SAbDab set with 90% training and 10% random test split. The optimized model achieved a training Root Mean Square Error (RMSE) of 1.70 on the log ($K_d$) scale. Feature importance analysis indicated that hydrogen bonds and hydrophobic surface area were the top predictors.

### D. AlphaFold2 Structure Prediction

For the hMPV A2.2 F protein, no experimentally determined structure was available. To generate a structural model, we used AlphaFold2 with the MMseqs2 pipeline to predict the 3D conformation using the input sequence retrieved from the NCBI Virus Variation Resource [17]. A phylogenetic tree constructed from these strains is shown in Fig. 7 (left), confirming the prevalence of sublineages A2.2.1 and A2.2.2 in recent outbreaks [18]. From this tree, a representative A2.2.2 F protein sequence was selected and submitted to AlphaFold2 for modeling.

The resulting structure showed strong internal confidence: the mean predicted Local Distance Difference Test (pLDDT) score exceeded 80 across conserved epitope regions, and the predicted surface geometry closely matched known structures, with over 80% sequence identity to previously characterized A2.2 variants. The 3D model (Fig. 7, right) highlights conserved epitopes in tan and key mutation sites, G42V and E96K, in pink.

To verify the model's reliability for downstream scoring, we analyzed contact maps across AlphaFold2's top-ranked predictions. As shown in Fig. 8, long-range contact density increased with rank, indicating folding convergence and structural robustness suitable for binding evaluation.

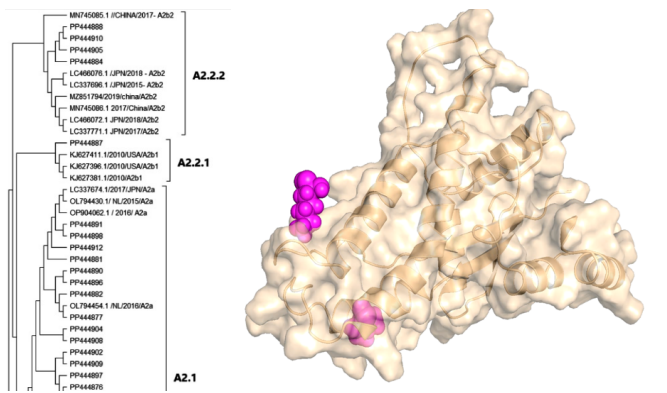

Fig. 7. AlphaFold2-predicted structure of the hMPV A2.2 F protein. Left: Phylogenetic tree of hMPV lineages used to identify the A2.2.2 variant [18]. Right: 3D structure showing conserved epitope regions (tan) and mutation sites G42V, E96K (pink).

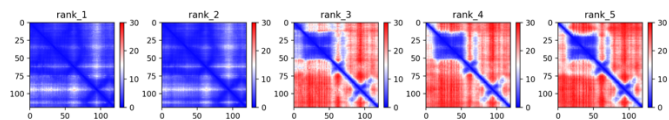

Fig. 8. Predicted contact maps for AlphaFold2 model ranks 1–5. Axes represent residue positions; color scale indicates residue-residue distance (Å). Denser long-range contacts suggest improved folding and reliability.

This pipeline enabled the use of predicted structures instead of missing crystallographic data.

*E. Transfer Learning: COVID-19 and hMPV Fine-Tuning*

To test the model's generalizability and adapt it to more relevant data, we applied transfer learning in two stages. First, the baseline model was fine-tuned on a dataset of 117 SARS-CoV-2 neutralizing antibody–antigen complexes (Ab-CoV dataset). This additional data (with $K_d$ labels) was used to fine-tune the last boosting iterations, yielding a final COVID-fine-tuned model. This process reduced the validation RMSE from 1.70 to 0.92 (a 46% decrease) as shown in Fig. 9.

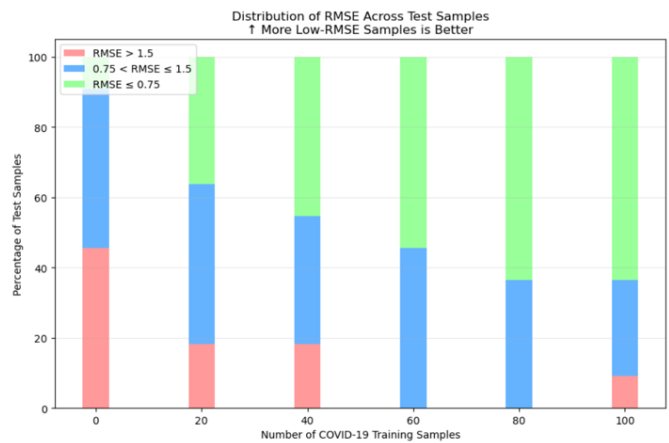

Fig. 9. Stacked bar chart showing the distribution of RMSE values across test samples as the number of COVID-19 training samples increases. Low-error predictions (green: RMSE ≤ 0.75) become more frequent, while high-error predictions (red: RMSE > 1.5) decline.

In the second stage, we performed leave-one-out cross-validation on the SARS-CoV-2 data to ensure stability, and then fine-tuned on known hMPV-F binders. Specifically, 22 hMPV F-specific antibodies (from SAbDab or literature) were used to update the model. The final hMPV-adapted model was then used to score a library of candidate antibodies (screened from virus-focused repositories and cross-reactive candidates) against the AlphaFold2 structure of hMPV A2.2.

## IV. RESULTS

*A. Baseline Model Performance*

The initial LightGBM model (trained on SAbDab) showed promising predictive power using only 3D structural features. On the held-out test set, the model achieved a median rank of 24 out of 213 for the true binder when sorting candidates by predicted affinity. This corresponds to an 89% reduction in search space: on average, one only needs to test 10% of candidates to find a high-affinity binder. At a stringent cutoff (top-10 predictions), the sensitivity was 30%, indicating true binders were often prioritized among the top few.

*B. COVID-19 Fine-Tuning*

Transfer learning on COVID-19 antibody data significantly improved the model. After fine-tuning on 117 SARS-CoV-2 complexes, the validation RMSE dropped by 46% (from 1.70 to 0.92 in log ($K_d$) units).

The violin plot (Fig. 10) demonstrates that the model's prediction confidence increased (with more predictions near true values). Regarding discrete outcomes, the proportion of predictions with error <0.5 log units rose from 60% to 80%. Feature importance stayed consistent post-fine-tuning, with hydrogen bonds and hydrophobicity as top predictors, confirming that the model captured key biophysical principles. The results show the framework can adapt to new viral contexts without full retraining.

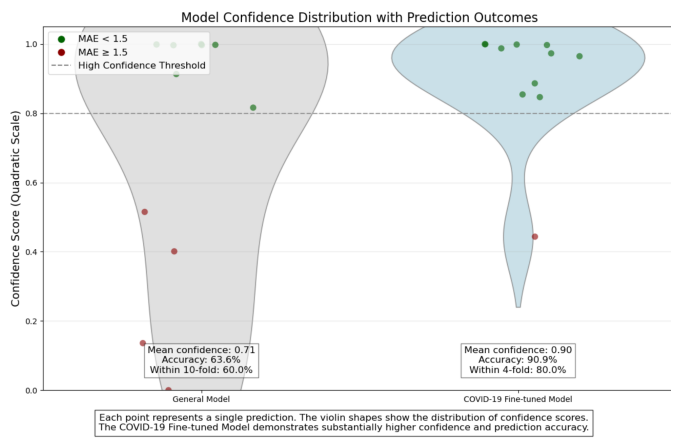

Fig. 10. Violin plot comparing prediction confidence distributions between the baseline LightGBM model and the COVID-19 fine-tuned model. Each dot represents a test prediction, colored by mean absolute error (MAE). The fine-tuned model shows higher mean confidence (0.90 vs. 0.71) and greater prediction accuracy (90.9% vs. 63.6%), with more predictions surpassing the high-confidence threshold.

## C. hMPV Antibody Predictions

Finally, the hMPV-adapted model was applied to screen candidate antibodies against the predicted hMPV A2.2 F protein structure. Two antibodies emerged as top hits: 6W16_HL [20] and 8E2U_HL [21]. These complexes were further analyzed structurally. The predicted binding affinities were in the picomolar range: $K_d = 5.0 \times 10^{-11}$ M for 6W16, and $1.1 \times 10^{-11}$ M for 8E2U.

Fig. 11 shows the docking poses: 6W16 binds near the mutated residue G42V, whereas 8E2U engages a conserved region away from the mutation. Both antibodies ranked in the top 2% of screened candidates. Interface analysis revealed significant surface complementarity between paratope and epitope with more than 6 hydrogen bonds in each complex, primarily involving complementarity-determining region (CDR) loops, and extensive hydrophobic contacts.

These findings suggest complementary binding strategies: one targeting the antigenic drift (G42V/E96K) and the other binding invariant regions. Importantly, these candidates had not been previously reported against hMPV A2.2, underscoring ImmunoAI's ability to generate valid hypotheses for experimental testing.

## V. DISCUSSION

The results validate the hypothesis that 3D structural descriptors enable rapid and accurate affinity prediction. The high dimensionality reduction (89% search space) indicates that this workflow can filter out most non-binders rapidly. The success of transfer learning on SARS-CoV-2 data shows that the model generalized beyond the original training set. Notably, fine-tuning improved accuracy and confidence (distribution of errors sharpened).

Our feature analysis, as shown in Fig. 12, aligns with established immunological principles. Hydrogen bonding and hydrophobic interactions emerged as the most predictive features. Lo Conte et al. found that protein–protein interfaces bury 1600 Å² of surface area, with interface atoms being close-packed and enriched in hydrophobic residues; a structural basis consistent with hydrophobic driving forces in binding [15]. The model's emphasis on interface SASA and polarity reflects core principles of paratope–epitope complementarity (mirroring observations from affinity maturation experiments).

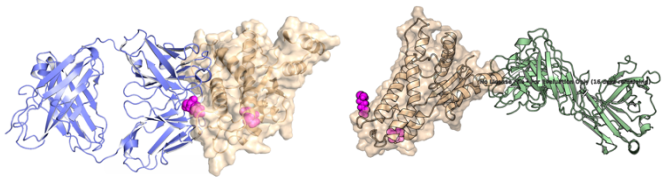

Fig. 11. Predicted binding of antibodies to hMPV A2.2 F protein. (A) Antibody 6W16_HL (blue) bound to epitope including mutation G42V. (B) Antibody 8E2U_HL (green) bound to conserved epitope. Predicted $K_d$ values indicate picomolar affinity.

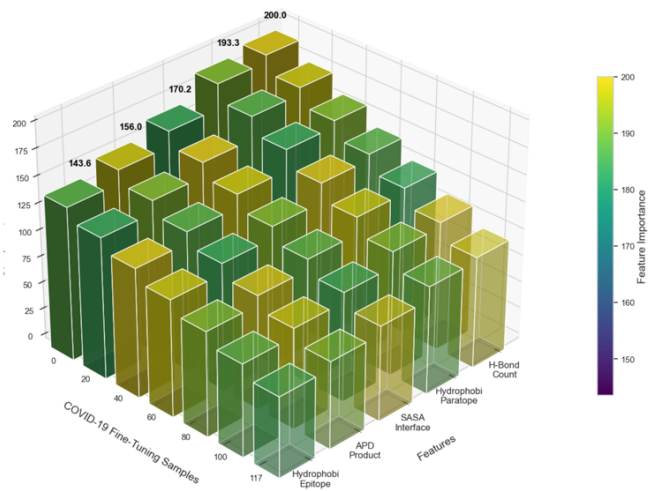

Fig. 12. Feature importance scores from the COVID-19 fine-tuned LightGBM model across structural descriptors. Hydrogen bonds and hydrophobicity consistently rank highest across sample sizes.

There are limitations to note. Although AlphaFold2 predicted the antigen structure with high confidence (pLDDT > 0.8, Fig. 13), the model remains static and may not capture conformational nuances such as loop mobility or induced-fit dynamics. Another limitation is that our study remains purely computational with a modest training set size: no experimental assays were performed here to validate the predicted antibody affinities, and results should therefore be interpreted as testable hypotheses rather than confirmed outcomes.

Future work will incorporate dynamic structural refinements using molecular simulations. As shown in Fig. 14, molecular dynamics, kinetic Monte Carlo, and diffusion-based methods can account for conformational uncertainty and flexible protein behavior across different timescales. Additional feature families such as evolutionary conservation, electrostatics, and developability scoring (e.g., aggregation propensity) also represent promising avenues to enhance candidate filtering and improve in silico antibody screening.

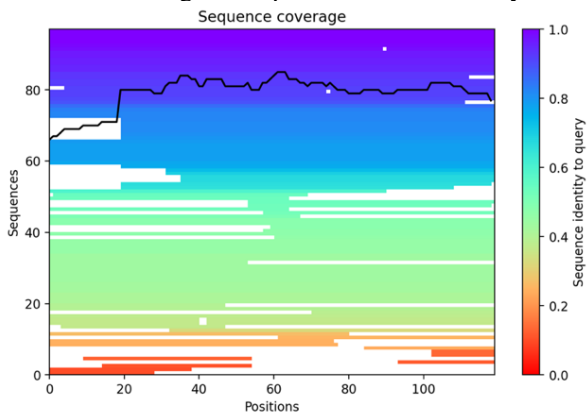

Fig. 13. Sequence identity heatmap for 89 hMPV F protein variants aligned to the A2.2 query. Colors indicate residue-level identity; the black line shows coverage depth. Conserved regions informed epitope selection for structural modeling.

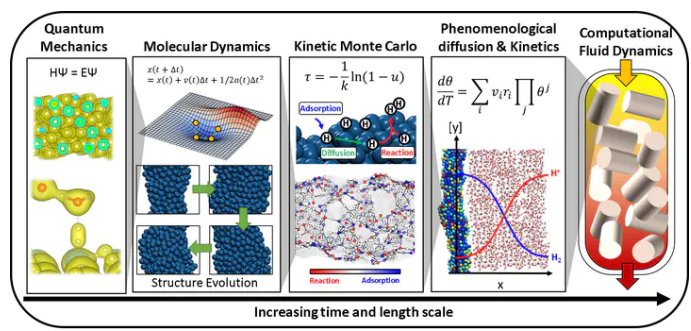

Fig. 14. Overview of dynamic simulation methods under conformational uncertainty for flexible protein modeling and future direction [11]

Despite these limitations, ImmunoAI's predictions offer a rational starting point for rapid antibody discovery and repurposing. Compared to deep-learning approaches, it offered a strong balance of performance and interpretability. Its success here is consistent with McElfresh et al. (2023), who reported tree-based models outperforming deep nets on antibody datasets under 1000 examples. Our model, despite not using explicit physics-based force fields, captured the relevant biophysical signals, suggesting that carefully chosen structural features alone carry sufficient information for affinity prediction.

The two hMPV A2.2 F protein antibodies identified here target biologically plausible epitopes with strong predicted interactions, but experimental binding assays will ultimately be required to validate these in silico results. If confirmed, this pipeline demonstrates a viable path to rapidly triage and propose therapeutic antibodies when facing novel viral variants during outbreaks where speed and timing are crucial.

## VI. Conclusion

We have presented ImmunoAI, a structural machine-learning pipeline for accelerated antibody discovery. By combining 3D interface feature extraction with gradient-boosted modeling, ImmunoAI significantly reduces the experimental search space and yields testable predictions of high-affinity antibodies. On benchmark data, ImmunoAI achieved 89% search space reduction. Transfer learning on SARS-CoV-2 data improved prediction accuracy. Using AlphaFold2-predicted antigen structures, ImmunoAI identified two candidate antibodies against the emergent hMPV A2.2 variant, with predicted affinities in the picomolar range.

In summary, ImmunoAI demonstrates that informed integration of structural bioinformatics and machine learning can dramatically speed up therapeutic antibody discovery. Future work will extend ImmunoAI to other pathogens (e.g., influenza, dengue) and incorporate molecular dynamics and developability filters, ultimately contributing to agile pandemic response toolkits.

## Acknowledgments

The authors would like to thank Professor Jyotishman Dasgupta and Dr. Ravi Venkatramani of the Tata Institute of Fundamental Research for their valuable guidance in shaping this project's methodology and scientific direction. The authors acknowledge Gokul Kannan (PhD student, Stanford University) for his insights and assistance.